\documentstyle[12pt]{article}
\begin{document}
\title{ Solutions of a 
Dirac hydrogenlike meson and 
scalar confinement at low orbital
angular momentum states. }
\author{
M. A. Avila \thanks{Electronic address: manuel@servm.fc.uaem.mx}
\\
  Facultad de Ciencias, UAEM, \\
      Cuernavaca 62210, Morelos, Mexico.\\  }

\date{}

\maketitle
\begin{abstract}

Wavefunctions of a heavy-light quark $(Q,\bar q)$
system described by a covariant Dirac hamiltonian 
are analyzed. 
By assuming that 
the confinement
potential is a
Lorentz scalar ($S$), 
the 
slope of the Isgur-Wise function 
is calculated
at zero recoil point. The result obtained is 
$\,\,\xi ^ \prime (1)\,=\, -0.93\,\pm\,0.05.\,\,$
This means that the solutions are perfectly
consistent.
If relativistic corrections
in the light quark wave functions
are included the result is
$\,\,
\xi ^ \prime (1)\,
=\, -1.01\,\pm\,0.04.\,\,$
From heavy-light data this suggests that
if relativistic effects are considered,
scalar confinement is reliable in low orbital
angular momentum states. 

\end{abstract}

\noindent{PACS number(s) : 
12.39.Pn; \, 12.38.Aw}

\newpage

\normalsize

A covariant Dirac-like hamiltonian 
describing a  
heavy quark-light quark 
$(Q,\bar{q})$
system 
has been introduced 
in Ref. \cite{avila}.
This hamiltonian 
contains all of the relativistic corrections. 
These effects come mainly from the recoil
of the heavy quark.
The model 
has been successful 
in applications involving both
hydrogenlike ($B-$,  $D-$) meson    
spectroscopy
and also in  
calculations of weak mixing angles \cite{solo}
from exclusive semileptonic B-decays. 
The hamiltonian of Ref. \cite{avila}
contains two 
potentials, one of them is  
a central linear Lorentz scalar potential ($S$)
which describes 
dynamically the confinement, and the other       
is a 
Coulomb potential ($U$) 
describing the short range color interquark interaction.
To be strong enough,  
the non-perturbative scalar potential $S$ does not feel
any relativistic corrections
but the Coulomb potential $U$
still is susceptible to them.
In fact, in the model of Ref. \cite{avila}
the potential $U$ incorporates 
relativistic corrections to first 
order in $m/M_Q$, where 
$m$ is the light quark mass
and $M_Q$ the heavy quark 
mass. 

One question
about the above description  
still remains to be studied 
and it is 
the concerning
to the 
consistency of its solutions.
For this reason, the purpose of this 
work is to examine the solutions
of \cite{avila} when the system
$(Q,\bar{q})$ is in low orbital
angular momentum states.
In order to do this, we 
subject the wavefunctions 
to
several  
phenomenological tests.
For instance, we check that they are
normalizable. The way in which this is
done is through 
an approach introduced in Ref. \cite{sucher}.
Another thing which is checked in
the present work 
is that the solutions account for 
the so called Klein paradox. This 
effect consists in the mixing 
of positive-negative energy 
states \cite{ovw}.  
The third and last test employed on the 
wavefunctions is to calculate the slope of the 
Isgur - Wise (IW) function at zero recoil
point, $\xi^\prime(1)$ \cite{hqet}. 
This quantity is calculated 
by assuming a scalar confinement
potential  
($S$) and also in the heavy
quark symmetry limit ($M_Q\,\rightarrow \, \infty$). 
The value obtained
for the slope is compared to other 
values of this quantity previously found
using different approaches.  
After we check that the wavefunctions 
of Ref. \cite{avila} are consistent,
then we include relativistic corrections. 
These corrected solutions 
are employed to make the corrections to 
$\xi^\prime(1)$ that result 
from a finite $M_Q$.  
Finally, the present work is concluded by 
discussing the nature 
of confinement in a $(Q,\bar{q})$
system in low angular momentum
states.

Let us first check the normalizability
of the solutions of Ref. \cite{avila}.
To do this, we follow 
an approach similar to that 
of Ref. \cite{sucher}. In \cite{sucher} 
it was pointed out that
a relativistic quarkonia system $(\bar {q},q)$
with a vector confining 
potential $\,\,V_v=\beta_1 \beta_2 \kappa_v r\,\,$
($\,\,\kappa_v>0\,\,$) 
could have 
unphysical 
non-normalizable solutions.
On the other hand, if  
the confining 
potential is of a Lorentz
scalar nature,
$\,\,V_s= \kappa_s r\,\,$ 
($\,\,\kappa_s>0\,\,$), 
this problem would not arise.
If one uses both kind
of potentials, 
it is necessary that scalar 
confinement be stronger than
vector confinement ({\it e.g.}
$k_s > k_v$)
in order to avoid 
singularities in the wave function.
In \cite{sucher} a possible 
singularity in the $(\bar q,q)$ 
wave function    
is avoided through the use of a      
Salpeter (no-pair) equation.

In order to illustrate 
the procedure, 
let us consider 
a simple scheme
where the $(Q,\bar{q})$ bound system 
is described by a Dirac equation. We 
assume that the 
only interaction existing between the quarks
is that of confinement and 
that the 
confinement
potential is composed of 
two parts. One of them is scalar, 
$ S = k_s r$ ($k_s>0$),
and the other vector, 
$V=k_v r$ ($k_s>0$). In 
this way, the respective equation
in the C.M. system is

\begin{eqnarray}
\Bigl[ {\bf \alpha} \cdot {\bf p} \,+\,m \beta\,+
\,\beta S\,+\,V \Bigr] \psi = E \psi,
\end{eqnarray}      

\noindent where ${\bf p}$ is the light quark momentum.

Let us decompose the wave function as 

\begin{eqnarray}
\psi\,=\,\psi^+ \,+ \,\psi^-, 
\,\,\,\,\,\,\,\,\,\,\,\,\,\psi^\pm \equiv P^\pm \,\psi, 
\end{eqnarray}      

\noindent where $P^\pm \equiv (1\, \pm \,\beta)/2$ 
are the 
standard projection operators into the 
upper and lower 
components of the wave function.
Then, by using (1) and (2), 
we obtain

\begin{eqnarray}
\psi^- \, = \, \frac 
{{\bf\alpha} \cdot {\bf p}}  {  E\,+\,m\,+ \, (k_s\,-\,k_v)\,r }
\,\,\,\psi^+.
\end{eqnarray}      

\noindent From the last expression
it is evident that if $E\,>\,0$ and 
$ k_v > k_s$, the lower sector of 
the wave function has a singularity
at $r \, = \,(E\,+\,m)/(k_v \, - \, k_s) $.
Consequently, the norm of the wave function

\begin{eqnarray}
<\,\psi \mid \psi\, >= 
<\,\psi^+ \mid \psi^+\, >
\, + \, <\,\psi^- \mid \psi^- \, >,
\end{eqnarray}      

\noindent is not
finite. 
In particular, we can observe  
from Eq. (3) that 
with a vector confinement ($S\,=\,0$), 
the wavefunctions
would not be finite.
But,   
if the confinement is scalar
($V\,=\,0$),
the solutions are finite.

Let us turn now to examine the normalization
of the wavefunctions
of the covariant   
hamiltonian which was introduced 
in  Ref. \cite{avila} 
for describing a 
$(Q,\bar{q})$ system.
This hamiltonian is

\begin{eqnarray}
\biggl[ {\bf \alpha}  \cdot 
{\bf p} \,\,+ \,\,
m \, \beta \,\, + \,\, \frac{p^2}{2 M_Q}
\,\, + \,\, M_Q \,\, + U(r) \,\, + &&
\nonumber   \\  
\nonumber
\\
\frac{U(r)}{2 \, M_Q}
\biggl({\bf \alpha} \cdot 
{\bf p} \,\,+\,\,
{\bf \alpha} \cdot { \bf \hat r} \,\, { \bf \hat r} 
\cdot {\bf p}
\biggr) \,\, + \,\, \beta S(r)\biggr]\,\psi\,
 = \,E\,\psi,
\end{eqnarray}      

\noindent where 
${\bf p}$ is the light quark momentum, 
$V_s(r) \equiv  \beta  S(r)=
\beta \kappa_s  r$
($\kappa_s>0$) is the 
scalar confining linear potential and 
$U(r)\,=\, - \xi / r$ ($\xi>0$)
is a color Coulomb-like 
potential. Note that 
terms proportional 
to $\frac{U(r)}{2 \, M_Q}$ and to
$\frac{p^2}{2\,M_Q}$ 
in this hamiltonian  
arise from 
the recoil of the heavy quark.

Using Eqs. (2) 
and (5) we obtain

\begin{eqnarray}
\psi^- & = &  
\frac{ 
\Biggl[\,\,\,\, {\bf \alpha} \cdot 
{\bf p} 
\,\, + \,\, \frac{U(r)}{2 \, M_Q} 
\biggl( \,\,\,
{\bf \alpha} \cdot {\bf p}\,\,+\,\, 
{\bf \alpha} \cdot {\hat{\bf r} }\,\,{\hat {\bf r}}\cdot{\bf p}
\,\,\,\biggr)\,\,\,\,
\Biggr]}
{E\,+\,m\,-\, \frac{p^2}{2\,M_Q}\,
- \, M_Q\, - \,U(r)\,+\,S(r)} \,\,\,
\psi^+,
\end{eqnarray}      

\noindent where $ \vert E \vert \,\geq \, M_Q\,+\, m $. 
Since 
$-\,U(r)\,+\,V(r)\,>\,0$ 
then for    
$ \epsilon\,\equiv\,E\,-\, M_Q\,-\, m\,\geq \, 0$ 
the wave function $\psi^-$
could  
have an unphysical singularity, which  
would depend strictly  
on the size of the term $\frac{p^2}{2\,M_Q}$ \footnote{In 
the infinitly heavy quark mass limit 
this term does not exist. Consequently 
if $\epsilon \,> \,0$, the solutions of Eq. (5) are
perfectly well defined for any value of $r$.}.
According to 
Heavy Quark Effective Theory (HQET) \cite{hqet}
this term represents the 
exchanged momentum
between the bound quarks.
It acquires 
a maximum value 
in the 
nonperturbative regime 
(quarks and gluons inside 
the meson) 
 \cite{neubert}. Thus, if we denote by 
$R$ the radius of the hadron, then
from HQET it follows that   
 
\begin{eqnarray}
\frac{p^2}{2\,M_Q} \,\sim\,1/R\,\,{fms}^{-1}
\,=\, 0.2/R\,\,GeV.
\end{eqnarray}      

\noindent Thus, if we define  
the denominator of Eq. (6) as

\begin{eqnarray}
h(r) \, \equiv \, 
{\epsilon \,+\,2m\,- \,U(r)\,+\,S(r)}
-\, \frac{p^2}{2\,M_Q},
\end{eqnarray}      

\noindent and use (7)
together with 
$\,\,\,U\,=-\,\xi/r\,\,fms^{-1}\,=\,-\,0.2\,\xi/\mid r \mid \,\,GeV,\,\,\,$
$\,\,\,S\,=\,\kappa_s\,r\,\,\,fms^{1}\,
=\, 5 \, \kappa_s \,\mid r \mid \,\, GeV^{-1},\,\,\,$
and $\,\,\, x\, \equiv \, r/R\,\,\,$, 
we obtain

\begin{eqnarray}
h(r) \, = \, 
{\epsilon \,+\,2m\,+ \,
\bigl( 
\kappa_s \,R^2 \, x^2\,- \, x
\, + \xi 
\bigr)/r }.
\end{eqnarray}

In order to show that $h(r)$
is positive for any value of $r$,
let us note that 
the quantity between parenthesis
in the last equation 
is positive at $\,\,x\,=\,0\,\,$
and it does not 
take negative values
for any value of $r$ 
providing that $\,4\, R^2\,\kappa_s\, \xi \,>\,1$.
For $\,\, \kappa\,=\,0.2\,GeV^{-2}\,\,$ 
and $\,\, \xi \, = \, 0.445 \,\,$ \cite{avila},
this last condition
means $\,\,\,  R \, > \, 1.6 \, GeV^{-1}
\,=\, 0.32\,fms.  \,\,\,$ Since this condition 
is very reasonable for a hadronic radius, 
we can 
conclude that $h(r)\,>\,0$ for
any value of $r$.
In particular, we have shown 
that the denominator $h(r)$
of Eq. (6)
is not zero for any value of $r$.
Therefore, 
the solutions of Eq. (5) are finite and
normalizable.

Let us proceed now with the second test
of the solutions of the hamiltonian
introduced in Ref. \cite{avila}. 
For this purpose
we impose the s-wave classical turning points  
condition ${\bf  p}\,=\,{\bf 0}$
in  
Eq. (5). Once we do this, 
we obtain $\beta \,(\,m\,+\,S\,)\,+\,M_Q\,+\,U\,=\,E$ 
which becomes

\begin{eqnarray}
\begin{array}{lll}
E_+ \,-\,M_Q\, =\,m\,+
\,S\,+\,U,
\end{array}
\end{eqnarray}

\begin{eqnarray}
\begin{array}{lll}
E_- \, -\,M_Q\, =\,-m\,-
\,S\,+\,U,
\end{array}
\end{eqnarray}      

\noindent for 
the positive ($\beta \,=\,1$) and 
negative ($\beta \,=\,-1$)
energy states respectively\footnote{ 
The condition ${\bf  p}\,=\,{\bf 0}$
can be seen from the 
MIT bag model point of view 
as the boundary condition
that prevents light quark $q$
current flux leaving through 
the meson bag.   
In this way, we expect 
typical values for the returning
point ($r^{r.p.}$) 
of order $\,\,
 r^{r.p.} \,  \geq  
\, R\,  \sim \, 0.78.\,\,fm\,\,$    
\cite{izat}-\cite{mit}}.

At this stage there are two 
possibilities,
either $U\,=\,0$
or $U\,\not=\,0$. If we turn off
the Coulomb interaction ($\,\,U \,=\,0\,\,$)
in the above equations we are  
reproducing the same analysis as Ref. \cite{ovw}. 
In this  work
the confinement 
in a Dirac equation was analyzed by assuming that the
confinement potentials were
of both kinds, scalar ($S$) and vector ($V$). 
It was found that the structure 
of the confinement must be 
scalar  
$(V=0)$ in order to avoid the 
Klein paradox (mixing of positive 
with negative energy states).
For the above reason, in the present 
work we are just considering 
the second and more realistic situation
where $U\,\not=\,0$.
Thus, if we turn  
on the color Coulomb interaction
and plot Eqs. (10) and (11)
with $\,\,\,S(r)\,=\,\kappa_s\,r\,\,\,$
and $\,\,\,U(r)\,=\,-\xi\,/\,r\,\,\,$
for two 
different values of the light quark masses
$\,m\,=\,0\,$, and $\,m\,=0.5\,\,GeV,$
we obtain the plot of Fig. 1.
As can be seen from this figure, 
the singularity in the Coulomb potential
at $\,\,r\,=\,0\,\,$ distorts and mixes the positive 
and negative energy states inside the  
domain of perturbative physics {\it i. e.}
$\,\,r\,\ll \, R \,\,$
\footnote{
In fact, as is well known, this mixing
inside the confinement 
region $r\,\ll\,R$ 
is responsible 
for the bound state spectra 
in a hydrogenlike
system $(Q,\bar{q})$. 
}. 
But (as should be expected 
of a good and consistent confinement potential)
this effect dissappears in  
the confinement  
region $\,\,r\,\simeq\,R$.
In other words,
Fig. 1 indicates that  
the scalar confining 
potential
$S$ of Eq. (5)
accounts for the Klein paradox
in the confining region $r\,\simeq\,R$
independent of the value of the 
light quark mass.
Furthermore,
we note also from Fig. 1 that  
the critical values of the
radial coordinate ($r_c$)
for which the physical
condition 
$\,\,E_+\,-\,M_Q\, - \,m
\, \geq \, 0$
begins to be valid, independently of the value 
of $m$, are those such that $\,\,r\,>\,r_c\,\,$ where 
$\,\,r_c\, \sim \,0.3 \,\,fms$. 
If we think
of 
$r_c$  
as the critical
value dividing
the perturbative 
and the non perturbative
region, it follows that   
the critical energy necessary to 
reach the non perturbative
region ($r\,>\,r_c$)
would be $\,\,
\epsilon_c\, \sim \, 1/r_c \, \sim \, 0.67\,\,GeV\,\,$.
Another sign of consistency is that precisely
$\,\,\,M_{D(D_s)}\,-\,M_c\, \sim \, 
M_B\,-\,M_b\,\sim\,0.66\,\,GeV\,\,\,\sim \epsilon_c$ \cite{avila}.

It is  
easy to see  
that a confining 
linear vector potential ($V\,=\,\kappa_v\,r$) 
without a scalar
potential ($S\,=\,0$) in Eq. (5)
is automatically discarded. In fact,  
in this case   
the s-wave returning point condition ${\bf p}\,=\,{\bf 0}$ 
would yield

\begin{eqnarray}
\begin{array}{lll}
 E_+ \,  -\,M_Q\,  = \,m+
\,V\,+\,U,
\end{array}
\end{eqnarray}

\begin{eqnarray}
\begin{array}{lll}
 E_- \, -\,M_Q\, =\, -m \,+
\,V\,+\,U.
\end{array}
\end{eqnarray}      

\noindent  
For large values of $r$
say $\,r \,\sim \,R$ where the  
Coulomb potential 
vanishes  
($U\,\sim \, 0$), then 
Eqs. (12) and 
(13) become    

\begin{eqnarray}
\begin{array}{lll}
\,E_-\,\simeq\,E_+\,\simeq\,M_Q\,+\,V\,>0\,;
\hskip1.50cm r \sim R.
\end{array}
\end{eqnarray}

\noindent This result is unphysical  
since it means that the negative and positive 
energy states 
are mixed in the non-perturbative 
region. In this way, we can conclude that 
only the scalar potential of Eq. (5)
can account for the Klein paradox. This result
makes the solutions
of (5) consistent.

In order to continue the analysis
of the solutions
of Eq. (5) let us turn
now to the calculation of the slope 
of the IW
function at zero recoil point.

By solving  
Eq. (5)
in the heavy quark symmetry limit  
($M_Q\,\rightarrow\,\infty$),
we find the light quark wavefunctions. 
With these solutions     
the IW function  
is calculated through the expression \cite{mit}

\begin{eqnarray}
\begin{array}{lll}
 \xi (\omega) \, = \,
\frac{2}{\omega \, + \, 1}
\,
\langle
\,
j_0(2\,E_q \sqrt{\frac
{\omega\,-\,1}{\omega\,+\,1}
} r)
\, \rangle,
\end{array}
\end{eqnarray} 

\noindent where $E_q$ is the light
quark energy. The average is defined as

\begin{eqnarray}
\begin{array}{lll}
 \langle
f
\rangle
 \, \equiv \,
\int_\Omega 
dr \,
r^2\,
\psi(r)^{\dag} 
\,
f(r)
\,
\psi(r)
,
\end{array}
\end{eqnarray}

\noindent where $\,\,\Omega\,\,$ is the spatial region 
explored by the wave function. 
In this work $\,\,\Omega\,\,$
is taken as a sphere
of radius $\,\,R\,\,$. 
Likewise, for the calculation
of the slope of the IW function at
zero recoil point, we use
both the solutions of (5) in
the heavy quark symmetry limit 
and the expression \cite{mit}

\begin{eqnarray}
\begin{array}{lll}
\xi^{\prime}(1) \, = \,
- \, \bigl(\,\frac{1}{2} \, + \, \frac{1}{3}\,E_q^2
<r^2>\bigr).
\end{array}
\end{eqnarray}

Once we do the above we obtain

\begin{eqnarray}
\begin{array}{lll}
\xi^{\prime}(1) \, \simeq \,
- 0.93\,\pm\,0.05.
\end{array}
\end{eqnarray}  

\noindent We have allowed the light quark mass 
to run from 
$0$ to $0.25\,\,\,GeV$ and the 
hadron radius 
as $R\,=\,0.78\,\,fm\,=\,3.9\,GeV^{-1}$ \cite{izat}.

The above value for $\xi^{\prime}(1)$ 
is consistent with values of this quantity 
previously obtained in other works.
For instance, in Ref. \cite{ovw} 
$\xi ^\prime (1)\, \simeq \, - \, 0.90$
was obtained by solving the Dirac equation
with scalar confinement ($S$)
through a variational method.
While in Ref. \cite{blok} by analyzing
sum rules
it was found $\xi ^\prime (1)\, \simeq \, - \, 0.65$.
In \cite{olsson}  
a relativistic flux tube mesonic
model was employed and gave the result   
$\xi ^\prime (1)\, \simeq \, - \, 0.93$, which is exactly
the same value found in the present
work.
The coincidences 
between 
(18) with values for $\xi ^\prime (1)$
obtained with such different 
approaches allow us to assert that solutions of 
(5) can go through this  
last test.

Since the solutions of (5) 
have passed satisfactorily 
the three different tests
above cited we can conclude that 
the model introduced in Ref. \cite{avila}
is perfectly consistent to describe
a $(Q,\bar q)$ system in low orbital
angular momentum states.

Now that we have checked 
the consistency
of Eq. (5), let us employ 
its solutions to discuss 
the nature of confinement in a 
$(Q,\bar q)$ system. 

As was pointed out
in Ref. \cite{ovw},
the value of the IW
function constitutes 
a sensitive test for 
the kind of confinement
potential.
According to this work, there is an apparent 
conflict at the moment between the 
values for $\xi ^\prime (1)$ obtained from a Dirac equation
with confining scalar on the one hand
and that 
extracted
from heavy-light `data' on the other.
The values of $\xi ^\prime (1)$
from the data are higher than that calculated
with a Dirac equation.
To quote just a few values obtained from different Lattice QCD (LQCD)
calculations,  $\xi ^\prime (1)_{LQCD}\,=\,-1.0$
\cite{bss}, $\xi ^\prime (1)_{LQCD}\,=\,-1.2$ 
\cite{booth}, and $\xi ^\prime (1)_{LQCD}\,=\,-1.16$.
Such a large discrepancy 
between $\xi ^\prime (1)$ 
and $\xi ^\prime (1)_{LQCD}$
in \cite{ovw} questions 
the scheme of a scalar 
confining potential in a Dirac equation
to describe a $(Q,\bar q)$ system.  
In fact, the authors of \cite{ovw} propose 
that if instead 
of a Dirac equation with scalar confining 
potential a Salpeter (no-pair) 
equation with a vector confining potential is used
to calculate
$\xi ^\prime (1)$, the
values obtained for $\xi ^\prime (1)$
would be in better agreement with 
heavy-light `data'.
The respective values for $\xi ^\prime (1)$
found in \cite{ovw} with these two different models 
are $\xi ^\prime (1)_{Dirac}\,\simeq\,-.90$
and $\xi ^\prime (1)_{no-pair}\,\simeq \,-1.2$.
From these values they then argue 
in favor of a no-pair equation 
with vector confinement.

We want to stress at this point that
the values obtained in
\cite{ovw} for $\xi ^\prime (1)$ 
do not contain in them
corrections 
coming from assuming $M_Q$ finite.
Furthermore, it is claimed in this work
that if one includes these corrections
in the light quark wave functions, 
the scheme of a Dirac equation with vector 
confining potential would be the correct one for the description
of a $(Q, \bar q)$ system. 
That these corrections 
are necessary is evident
from the following three facts:
1. The limit $M_Q\,\rightarrow \, \infty $
where both the IW function 
and its slope are properly 
defined are just a good approximation 
to reality. 
2. Even in the limit $M_Q\,\rightarrow \, \infty $, 
the IW function and its slope
are of a non-perturbative nature.
Consequently
their 
values rely on parameters (quark masses,
Bag radius, tensions, etc) independently
of the approach employed.
3. A good model for the description of a $(Q,\bar q)$ 
is that which retains its validity in the limit
$M_Q\,\rightarrow \, \infty $.
Furthermore, this model
can be seen as a good heavy quark
picture modified by finite
relativistic corrections.

As we stated above, we 
consider it important 
to include relativistic corrections to
the slope of the IW form 
factor at zero recoil point. 
The way we incorporate 
these relativistic
corrections to $\xi ^\prime (1)$
is as follows. 
First, we obtain
the light quark wavefunction 
from Eq. (5) with finite $M_Q$ using for this 
purpose the same parameters as in Ref. \cite{avila}.
These 
solutions are then substituted in Eq. (17).
Once the above procedure is implemented, we obtain 

\begin{equation}
\xi ^\prime (1)_{corr}\,=\,-1.01\,\pm\,0.04,
\end{equation} 

\noindent where the light quark mass
has taken the same range of values 
as in Eq. (18).

As can be seen in Eq. (19)
the value for $\xi ^\prime (1)$
with relativistic corrections 
is in considerably better agreement with 
heavy-light data than those
without relativistic corrections 
in Eq. (18) and Ref. \cite{ovw}.  
The corrected value
for $\xi ^\prime (1)$   
of Eq. (19) modifies sustantially any 
conclusion about the nature 
of confinement. 
In fact, this result 
induces us to conclude that 
a Dirac equation with scalar confinement 
constitutes a good way of describing  
a $(Q, \bar q)$ system 
in low orbital angular momentum states 
providing 
relativistic corrections are taken into account.

\newpage

{{\bf Acknowledgement} 
I thank to N. A. 
and to N. H. with whom part of this 
work was done. I thank also to M. S. for 
helping with the preparation of the manuscript.
I acknowledge support
from CONACyT grant 3135.}

\newpage
\vskip5.0cm 
\centerline{FIGURES}
 
\vskip3.0cm

\noindent Figure 1. Behavior of the right
hand side of Eqs. (10)
and (11) with $U(r)\,=\, - \xi / r$ and
$S(r)\,=\,\kappa_s\,r$ for two different
light quark masses: 1a). $m=0$
and 1b). $m\,=\,0.5\,\,\,\,GeV$.
The values employed for $\xi$
and $\kappa_s$ are the same of Ref. \cite{avila}
{\it e. g.}
$\xi\,=\,0.445$ and $\kappa_s\,=\,0.2\,\,\,GeV^{-2}$. 

\newpage

\begin{thebibliography}{9}

\bibitem{avila} M. Avila, Phys. Rev. {\bf D 49},  
(1994) 309. 


\bibitem{solo} M. Avila, J. Phys. {\bf G 21}, (1995) 615.


\bibitem{sucher} J. Sucher, Phys. Rev. {\bf A 22} 
(1980) 348;
Phys. Rev. {\bf D 51}, (1995) 5965.



\bibitem{ovw} M. G. Olsson, S. Veseli, and K. Williams,
Phys. Rev. {\bf D 51}, (1995) 5079.



\bibitem{hqet} N. Isgur and M. B. Wise, Phys. Lett. {\bf 69} (1989) 111 ;
		   {\it ibid}\,{\bf 237} (1990) 527.


\bibitem{neubert} M. Neubert, Phys. Rept. {\bf 245} (1994) 259.  


\bibitem{izat} D. Izat, C. De Tar and M. Stephenson,
Nucl. Phys. {\bf B \, 199} (1982) 269.


\bibitem{mit} M. Sadzikowski and K. Zalewski, 
Z. Phys. C {\bf 59} (1993) 677; H.
H$\not$ogaasen and M. Sadzikowski, Z. Phys. C {\bf 64}
(1994) 427.



\bibitem{blok} M. Blok and 
M. Shifman, Phys. Rev. {\bf D 47} 
(1993) 2949.


\bibitem{olsson} M. G. Olsson and S. Veseli, 
Phys. Rev. {\bf D 51}, (1995) 2224.


\bibitem{bss} C. W. Bernard, Y. Shen
and A. Soni,
Nucl. Phys. (Proc. Suppl.)
{\bf B 30}, (1993) 473.



\bibitem{booth} S. P. Booth et al, 
Phys. Rev. Lett. {\bf 72}, (1994) 462.



\end{thebibliography}
\end{document}